# Privacy-Enhanced Reputation-Feedback Methods to Reduce Feedback Extortion in Online Auctions

Michael T. Goodrich

Dept. of Computer Science
University of California, Irvine
Irvine, California 92697-3435 USA

Florian Kerschbaum

SAP Research
Vincenz-Priessnitz-Str. 1
76131 Karlsruhe, Germany

**Abstract**

In this paper, we study methods for improving the utility and privacy of reputation scores for online auctions, such as used in eBay, so as to reduce the effectiveness of feedback extortion. The main ideas behind our techniques are to use randomization and various schemes to escrow reputations scores until appropriate external events occur. Depending on the degree of utility and privacy needed, these external techniques could depend on the number and type of reputation scores collected. Moreover, if additional privacy protection is needed, then random sampling can be used with respect reputation scores in such a way that reputation aggregates remain useful, but individual reputation scores are probabilistically hidden from users. Finally, we show that if privacy is also desired with respect to the the reputation aggregator, then we can use zero-knowledge proofs for reputation comparisons.

## 1 Introduction

Now that online auctions are a mature Internet enterprise, online auctions are a common source of complaints of Internet fraud. The most common first line defense for dealing with auction fraud is to provide a feedback system, so that buyers and sellers can rate each other. That is, to provide a means for deterring cheating and fraud, the managers of online auction systems allow sellers to rate buyers on how quickly and honestly they pay for the items for which they are the highest bidder and buyers to rate sellers on how honestly they described their goods and how well they shipped them. Not surprisingly, the feedback rating score of a buyer or seller can have dramatic impacts on their ability to buy and sell goods [19].

Rating systems vary from those that rate on a numerical scale to those, such as eBay, that rate on a simple "negative," "neutral," and "positive" continuum. In practice, however, these different feedback mechanisms are similar, in that anything but the highest, most positive rating is seen as a bad score. Thus, we take the simplified view in this paper that feedback evaluations are essential binary, being either "positive" or "negative."

Because of the importance of reputation in online auctions, there is a strong incentive for buyers and sellers to work hard to get highly positive ratings. Naturally, the intent of this incentive is that buyers and sellers should be on their best behavior, with buyers being fast and reliable with their payments and sellers being fast and reliable with their item descriptions and shipping methods.

Unfortunately, dishonest or incompetent buyers and sellers still want positive feedback, and some are even willing to manipulate the system to get it. In particular, some buyers and sellers who deserve negative feedback will nevertheless pressure their trading partner for positive feedback even if their behavior is more accurately classified as "negative." For example, a seller might threaten to provide negative feedback to an unhappy buyer if she provides negative feedback on the seller.

Such *feedback extortion* became such a problem within eBay, for example, that in January 2008 eBay changed its feedback policy so that sellers can only provide positive feedback for buyers. In addition, the eBay web site contains the following feedback extortion policy[1]:

> Buyers aren't allowed to demand goods or services outside of the transaction while threatening negative Feedback, neutral Feedback, or low detailed seller ratings.
>
> Sellers can't require buyers to leave specific Feedback or detailed seller ratings. Sellers also can't demand that buyers withdraw existing Feedback or detailed seller ratings. This applies to all Feedback activity, whether it happened before, during, or after delivery of items or services described in the original listing.

This solution removes the possibility of feedback extortion from sellers, of course, but it also completely negates the usefulness of feedback on buyers. Now the only influence that a seller can provide on the reputation of a poor buyer is to say nothing at all and hope that other, external methods can be used to resolve any disputes with buyers. Even then, this seller feedback restriction approach still does not stop feedback extortion from buyers, who can still threaten negative feedback on sellers unless they give the buyers positive feedback by a certain deadline. Thus, we are interested in this paper on schemes that can promote good-quality, accurate feedback, while discouraging or even preventing feedback extortion.

## 1.1 Prior Related Results

The positive impact of feedback on interacting parties has been studied by economists for a long time and even been experimentally verified [4]. Nevertheless only the bilateral feedback system enabled the emerging problem of feedback retaliation.

### 1.1.1 Feedback Correction and Evaluation

Feedback retaliation is well-known folklore in online auction sites. Resnick and Zeckhauser systematically evaluated eBay feedback data [19]. They noticed that only 0.3 percent of all transactions were rated negative, but in case one partner rated negative the probability of the other partner rating negative was over 37%. Furthermore, only roughly half of the transactions were rated at all. This data at least suggests that there is a problem of feedback extortion and reluctance to rate negatively.

Miller et al. suggest a payment mechanism to elicit honest feedback [14]. Such a payment mechanism can be implemented cost-neutral to the reputation platform and have a Nash equilibrium for truth-telling. Of course, its realization involves another accounting mechanism not necessary in our implementation.

Traupman and Wilensky suggest a statistical method in order to correct the error from negative ratings [20]. Nevertheless they evaluate their system only under synthetic data from current reputation systems. Clearly this does not take into account the game changing effect a change in the reputation scoring algorithm has. For example, if it becomes less threatening for someone to be rated negative, he might be inclined to even lower his ratings or perform worse.

---

[1]`http://pages.ebay.com/help/policies/feedback-extortion.html`

### 1.1.2 Feedback Computation Privacy

Privacy as a mechanism to protect against feedback extortion has been proposed previously by Kerschbaum [12], in that he presents a protocol for a private and reliable reputation system using two mutually distrustful service providers, i.e., where there is no single trusted reputation provider anymore. Nevertheless, the proposed mechanism still suffers from a long delay before publishing results, which this paper removes via sampling and escrow.

Note that, in the context of this paper, privacy deals with the secrecy of a rating score, i.e., *how* an agent was rated. Other private reputation systems, e.g., [1,15,18], only protect the identity of the rater, i.e., *who* has rated, which does not protect against feedback extortion.

Privately computing a reputation score is an instance of secure computation (SC) [2,11,21]. SC allows a number of parties to compute a function (such as a reputation score) on joint inputs (such as the ratings) without disclosing anything except what can be inferred from one party's input and output. We stress that a straight-forward application of SC does not protect the rating by itself. The continuous release of the reputation score, i.e., the result of the aggregation mechanism, reveals the input and breaks privacy. Protocols that implement secure computation therefore either use a more complicated reputation mechanism, e.g. collaborative filtering [5], or simply do not reveal the result [17].

### 1.1.3 Differential Privacy

The work of this paper is also related to approaches used in the context of *differential privacy* (e.g., see [9]) for hiding the contribution of a single value to the whole through the introduction of random noise or assumed randomness. Unfortunately, such approaches are not a good fit for protecting the privacy of reputation scores in online auctions. There are two primary reasons for this difficulty.

First, most existing schemes, such as that by Dinur and Nissim [7], for protecting the privacy of individual values that contribute to a sum require that one at least knows the number of items included in the sum. But reputation scores have an arbitrary, unbounded number of items that can be included—they are the total number of items bought or sold by a member of an auction—and making such previous differential-privacy methods work for such unbounded summations appears difficult.

Second, a reputation score evolves over time. So we cannot assume that such scores are drawn uniformly at random from some distribution, say, as is needed in the differential-privacy scheme of Duan [8]. Likewise, a participant that queries his or her own reputation after every transaction can be modeled as issuing correlated summation queries, but existing techniques, such as that of Li *et al.* [13], for obfuscating such queries, appear to introduce too much noise to be of practical use for reputation aggregation.

## 1.2 Our Results

In this paper, we first present a game-theoretic analysis of simple reputation escrow. In simple reputation escrow we consider a single transaction and its feedback. We show that when players are forced to leave feedback concurrently they are inclined to leave correct feedback. This applies even in case both players experienced a negative transaction.

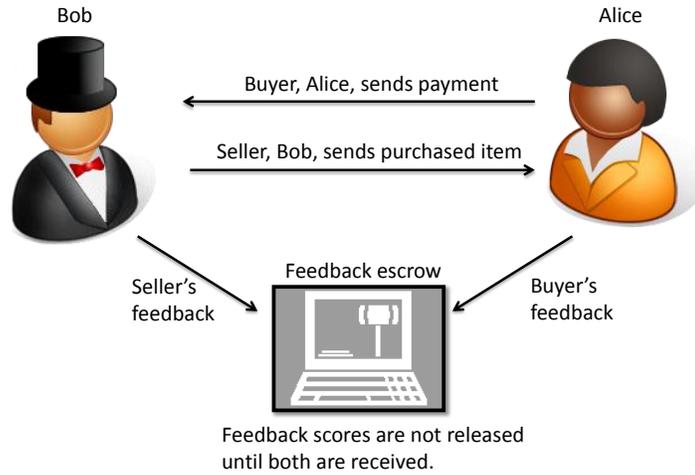

Figure 1: Feedback escrow.

We then extend this result to a reputation system that keeps feedback escrowed forever, i.e., it remains entirely private. This may seem impossible at first sight, since changes in the reputation score reveal the type of left feedback. Nevertheless we introduce a randomized sampling technique that can compute reliable averages while still providing privacy of the feedback. We show the error bounds of this sampling technique and conclude that it is a reliable estimator of performance.

As a last result, we remove the need for a trusted reputation service provider. We adapt the previous result of Kerschbaum [12] to include the randomized sampling technique. This adaption requires an additional zero-knowledge proof for correct sampling. The challenge in its construction is that the input for the proof is distributed.

In summary, this paper contributes

- a *game-theoretic analysis of reputation escrow*

- a *randomized feedback sampling mechanisms* that provides privacy of ratings despite immediate publishing

- a *secure computation of a zero-knowledge proof* for correct sampling that removes the need for a trusted reputation service provider

The remainder of the paper is structured as follows: Section 2 introduces reputation escrow and presents its game-theoretic analysis. Section 3 describes the sampling algorithm and analyses its error bounds. Section 4 explains how to remove the need for a trusted reputation provider and the secure computation protocol for the necessary zero-knowledge proof including a security proof. We conclude in Section 5.

# 2 Simple Reputation Escrow

The first solution we consider is a simple one—escrow the feedback from a buyer or seller, which ever comes first, until the second party in the transaction posts feedback on the first. (See Figure 1.)

This simple approach removes the sequential nature of feedback scoring for determining a person's reputation in the online auction; hence, it prevents quid pro quo retaliation or reward, for non-repeated transactions. When an online auction uses a reputation escrow, a buyer or seller cannot know the feedback score they are receiving from the other until they post their own feedback score for the other. Rather than argue the benefits of a reputation escrow in an ad hoc manner, however, let us analyze its benefits using game theory.

## 2.1 Studying Reputation Escrow Through a Game Theoretic Lens

When a buyer, Alice, and seller, Bob, are contemplating the feedback to give one another after a transaction, there are several factors that impact their respective decisions. Let us model what we feel are the most important of these factors using the following non-negative parameters (where we let $x$ represent $A$ for "Alice" or $B$ for "Bob" and we let $y$ denote the other party, i.e., "Bob" or "Alice")

- $\Delta_x$: the utility that $x$ receives by having his or her reputation factor increase from the receipt of a positive feedback score. So $-\Delta_x$ is the amount by which $x$'s reputation utility would decrease from his/her receiving a negative feedback score.

- $f_x$: the normalized utility that $x$ receives from giving a correct feedback (e.g., from the satisfaction of doing the right thing and/or helping other auction users learn the true reputation of $y$ better).

- $r_x$: the normalized utility that $x$ receives from performing a revenge punishment on $y$ for a negative feedback from $y$. It is a recognition of human nature to allow for this parameter, but we nevertheless assume that revenge satisfaction cannot fully overcome negative feedback, even when one is also giving out the correct feedback (so $r_x + f_x < \Delta_x$).

There are four different games that Alice and Bob can play, depending on whether each wants to give the other a positive (P) feedback or a negative feedback (N). We can model these games using a two-by-two payoff matrix, where we let Alice be the row player and let Bob be the column player. These are shown in Figure 2 in generic form.

If Alice and Bob both want to give positive feedback (the P-P game), then this is either a game strictly dominated by the (P,P) response (if $r_x < f_x$), which is the unique Nash equilibrium in this case, or (if $r_x > f_x$) it is an instance of a generic stag hunt game[2]. In the case of a stag hunt game (i.e., if $r_x > f_x$), there two Nash equilibria for this game, the (P,P) response and the (N,N) response. Clearly, however, the (P,P) response is preferred by both parties, even in this case, so it is the most likely response, as one would expect, even if $r_x > f_x$.

The N-P and P-N games are less interesting, from a game theoretic point of view, in that one player has a dominating N strategy and, given that choice, the other player's best choice is either

---

[2] In a *stag hunt* game, two players are hunting a stag. If they cooperate, they both share the prize. If one of them does not cooperate, while the other still does, then the non-cooperator gets a hare (which is worth less than half a stag), while the stag hunter gets nothing. If they both defect, they both get hares.

|   | P | N |
|---|---|---|
| P | $(\Delta_A + f_A, \Delta_B + f_B)$ | $(-\Delta_A + f_A, \Delta_B)$ |
| N | $(\Delta_A, -\Delta_B + f_B)$ | $(-\Delta_A + r_A, -\Delta_B + r_B)$ |

(a)

|   | P | N |
|---|---|---|
| P | $(\Delta_A + f_A, \Delta_B)$ | $(-\Delta_A + f_A, \Delta_B + f_B)$ |
| N | $(\Delta_A, -\Delta_B)$ | $(-\Delta_A + r_A, -\Delta_B + r_B + f_B)$ |

(b)

|   | P | N |
|---|---|---|
| P | $(\Delta_A, \Delta_B + f_B)$ | $(-\Delta_A, \Delta_B)$ |
| N | $(\Delta_A + f_A, -\Delta_B + f_B)$ | $(-\Delta_A + f_A + r_A, -\Delta_B + r_B)$ |

(c)

|   | P | N |
|---|---|---|
| P | $(\Delta_A, \Delta_B)$ | $(-\Delta_A, \Delta_B + f_B)$ |
| N | $(\Delta_A + f_A, -\Delta_B)$ | $(-\Delta_A + f_A + r_A, -\Delta_B + f_B + r_B)$ |

(d)

Figure 2: The four games that come from Alice and Bob respectively wanting to give (a) P-P feedback, (b) P-N feedback, (c) N-P feedback, and (d) N-N feedback.

P or N depending on the relative utility they receive from altruism (their $f_x$ value) or revenge (their $r_x$ value). So, if $r_x > f_x$, then (N,N) is the unique Nash equilibrium for both of these games. Alternatively, if $r_x < f_x$, then (N,P) or (P,N) is the unique Nash equilibrium, depending on whether it is Alice or Bob that feels the other deserves a negative feedback (i.e., whether they are playing the N-P game or the P-N game).

The N-N game, on the other hand, is an instance of a generic prisoner's dilemma game[3]. In spite of the dilemma posed by the fact that (P,P) is a highest payoff choice for both players, the Nash equilibrium for this game is (N,N).

So as to provide more concrete examples of these games, we show examples of these games and in Figure 3 using the concrete values, $\Delta_x = 5$, $f_x = 3$, and $r_x = 1$. Note that in this case, where revenge is less powerful than altruism, i.e., when $r_x < f_x$, then we get that the Nash equilibrium for each game is the desired feedback that we would like to see for this reputation system. That is, the system works in this case, in that the optimal choice for each player in this system is to provide the feedback that they feel is most appropriate for the other player.

Of course, even in the case when revenge is stronger than altruism, the two parties don't know with certainty which game they are actually playing, between the games P-P, P-N, N-P, and N-N. At best, they can eliminate two of these games, based on what they think is the appropriate feedback to give the other person. But they don't know the feedback the other person feels is most

---

[3] In a *prisoner's dilemma* game, two prisoner's are accused of a crime. If they both deny the other is guilty, then they both go free. If only one testifies against the other, on the other hand, then he gets a reward and goes free, while the other gets a heavy sentence. But if both of them testifies against the other, they both get modest prison terms.

|   | P | N |
|---|---|---|
| P | (8, 8) | (−2, 5) |
| N | (5, −2) | (−4, −4) |

(a)

|   | P | N |
|---|---|---|
| P | (8, 5) | (−2, 8) |
| N | (5, −5) | (−4, −1) |

(b)

|   | P | N |
|---|---|---|
| P | (5, 8) | (−5, 5) |
| N | (8, −2) | (−1, −4) |

(c)

|   | P | N |
|---|---|---|
| P | (5, 5) | (−5, 8) |
| N | (8, −5) | (−1, −1) |

(d)

Figure 3: The four games that come from Alice and Bob respectively wanting to give (a) P-P feedback, (b) P-N feedback, (c) N-P feedback, and (d) N-N feedback, for the concrete values, $\Delta_x = 5$, $f_x = 3$, and $r_x = 1$.

appropriate to give them. Based on the above analysis, if a person feels the other player deserves a negative feedback (N), then they should give them a negative feedback. But, even if revenge is stronger than altruism (i.e., $r_x > f_x$), a player who feels the other person deserves a positive rating (P), has to make an educated guess as to whether they are playing the P-P game or the N-P or P-N game (depending on whether they are the buyer or seller). By assigning probabilities, based on how satisfied they think the other person is, a player can calculate an expected return based on their probabilities that they are playing, say, the P-P game or the P-N game. In most cases, the optimal choice in this instance will be to provide positive feedback for the other person when they deserve it. Thus, even if revenge is stronger than altruism, the reputation system should work and it should score people with their appropriate feedback scores most of the time.

## 2.2 Why Non-Escrowed Systems Breakdown

In a nutshell, a crucial property that allows the reputation escrow system to work is that the buyer and seller have a degree of uncertainty about what the other person is going to do. In fact, they don't know for certain even what game they are playing, between the choices of P-P, N-P, P-N, and N-N. And that uncertainty pushes each person to provide the appropriate level of feedback as their optimal response. In this subsection, we outline why removing this uncertainty, as in a non-escrowed system, allows for feedback extortion.

Suppose now that the feedback game is played sequentially, so that player $x$ first makes their choice, which is revealed to player $y$, and then $y$ makes their choice. For simplicity, let us assume that it is the buyer, Alice, who goes first and the seller, Bob, who goes second (the reverse scenario is similar). Notice immediately that as soon as Alice provides her feedback, Bob knows her response and can react to that choice with certainty. Moreover, Alice now knows that Bob is going to realize her response as soon as she reveals it, and this can in fact influence the choice Alice is going to make.

This influence doesn't make much of a difference, however, in three of the feedback games. In the P-P game, for instance, Alice reveals a positive response and Bob replies with a positive response, as these are the optimal choices for them. Likewise, in the P-N game, Alice reveals her positive or negative response depending on whether she favors altruism over revenge, just as in the synchronized version, and then is hit with a negative feedback from Bob either way. And in the N-N game, Alice reveals a negative response and is immediately hit with negative feedback from Bob, as this is the equilibrium choice for both of them.

Something interesting happens in the N-P game, however, where Alice feels that the correct assessment of her satisfaction with the transaction is negative while Bob feels that the correct assessment of his satisfaction with the transaction is positive. Let's looks at the possible scenarios:

- *Alice reveals a positive response.* In this case, Bob's optimal response is also positive, and the payoff is $(\Delta_A, \Delta_B + f_B)$.

- *Alice reveals a negative response.* In this case, depending on Bob's relative utility between altruism and revenge, his optimal response is either positive, with payoff $(\Delta_A + f_A, -\Delta_B + f_B)$, or negative, with payoff $(-\Delta_A + f_A, -\Delta_B + r_B)$.

Thus, Alice knows that if she reveals a positive response, she is giving up on a higher payoff, of $\Delta_A + f_A$, which she might get from Bob should he be an altruistic person. But this is exactly the decision that Bob can now influence. Should he reveal to Alice ahead of her choice (e.g., using an email threat or even just a snide comment) that he prefers revenge over altruism, then her optimal choice is now to give him positive feedback. That is, she is making a suboptimal choice for herself, based on pressure from Bob, which is extortion. Of course, Bob would not know at the point of his threats whether Alice and he are playing the P-P, N-P, P-N, or N-N game, but he loses nothing in the P-P, P-N, or N-N game from revealing his preference for revenge over altruism. Since he gains something in the N-P game from this revelation, Bob is therefore motivated to threaten Alice with retaliation should she give him negative feedback. The reputation escrow system reduces the possibility of this abuse, however, by making it impossible for Bob to know with certainty the feedback he is getting from Alice (and vice versa) until after he reveals his feedback score.

Of course, if Bob strongly suspects that Alice has given him negative feedback, then he may refuse to give Alice any feedback, so as to avoid getting her feedback. So this simple reputation escrow approach has the drawback that it allows for Bob to effectively block feedback that he anticipates will be negative. The simple reputation escrow system returns the feedback process to have the players giving their honest feedback when they are both motivated to complete the feedback process. But it doesn't stop this feedback blockage that either Alice or Bob could initiate against the other. The next solution we describe addresses this drawback.

# 3 Sampling from Archived Reputation Scores

Let us now consider a more restrictive reputation escrow system, where we escrow all feedback and never release it in its raw form, ever. That is, let us consider an *archived* reputation escrow. At first, this might seem to be a wasted effort, which could never yield useful reputation data, but there is a work around.

Since one of the main reasons for having a reputation feedback score is so that we can compute reputation averages for each participant in an online auction, let us consider how we could use an archived reputation escrow to compute a reputation score.

Suppose that Alice has just left feedback for Bob, and, like all of his feedback scores, this score is not released. We nevertheless would like to compute a reputation score, which is the average of his reputation scores (e.g., on a scale of 1 to 5, for degree of satisfaction, 0 to 1, for positive-negative, or $-1$ to 1, for negative-neutral-positive). Let $n$ denote the number of feedback scores that Bob now has and let $S$ denote the sum of these scores. Thus, we are interested in computing $S/n$, Bob's average feedback score. But if we directly release $S/n$ and Bob has been keeping track of how this average score changes each time he receives a new feedback score (and, in particular, Bob knows this average before Alice gives her feedback), then Bob can figure out how Alice rated him.

So instead of directly computing $S/n$, we independently sample from Bob's feedback scores $r < n$ times, with each score being equally likely (and allowing for repetitions, since each selection is independent). We then compute the sum, $T$, of the chosen scores and we release Bob's computed average feedback score, $F_B$, as

$$F_B = \frac{Tn}{r},$$

where the parameter $r$ is determined below in our analysis. So let us study the degree of accuracy and privacy that comes from this sampling approach.

## 3.1 Accuracy

Let us simplify our analysis by assuming that the scores are either 1, for "positive," or 0, for "negative." This is actually not much of a restriction in practice, since praise inflation in online auctions occurs to such a regular degree that anything but the highest feedback score is considered negative. That is, buyers tend to rate sellers by the percentage of feedbacks that are the highest possible.

Let $X_i$ be a random variable that is 1 if the $i$th feedback score chosen from Bob's feedbacks is positive, so $X_i = 0$ if this score is negative. So we can write

$$T = \sum_{i=1}^{r} X_i.$$

To analyze the accuracy of this random sampling method, let us consider the expected error that occurs from this approach,

$$E\left(\left|\frac{T}{r} - \frac{p}{n}\right|\right),$$

where $p$ is the number of positive feedbacks that Bob has.

For relatively small values of $n$ and $r$, we can compute this expected value directly as

$$\sum_{i=0}^{r} \left|\frac{T}{r} - \frac{p}{n}\right| \Pr(T = i),$$

which is the same as

$$\sum_{i=0}^{r} \left|\frac{i}{r} - \frac{p}{n}\right| \binom{r}{i} \left(\frac{p}{n}\right)^i \left(1 - \frac{p}{n}\right)^{r-i}.$$

The expected error is maximized for $p = \frac{n}{2}$. This implies that the necessary sample size $r$ for achieving an expected error of 5% or 10%, respectively, is constant and independent of the number of feedback scores $n$. Figure 4 depicts the expected error for sample sizes of 1 to 75. We conclude that 16 samples are sufficient to lower the expected error below 10% and 64 samples for 5%. Consequently it is sufficient and efficient to sample a (small) fixed number of feedback scores.

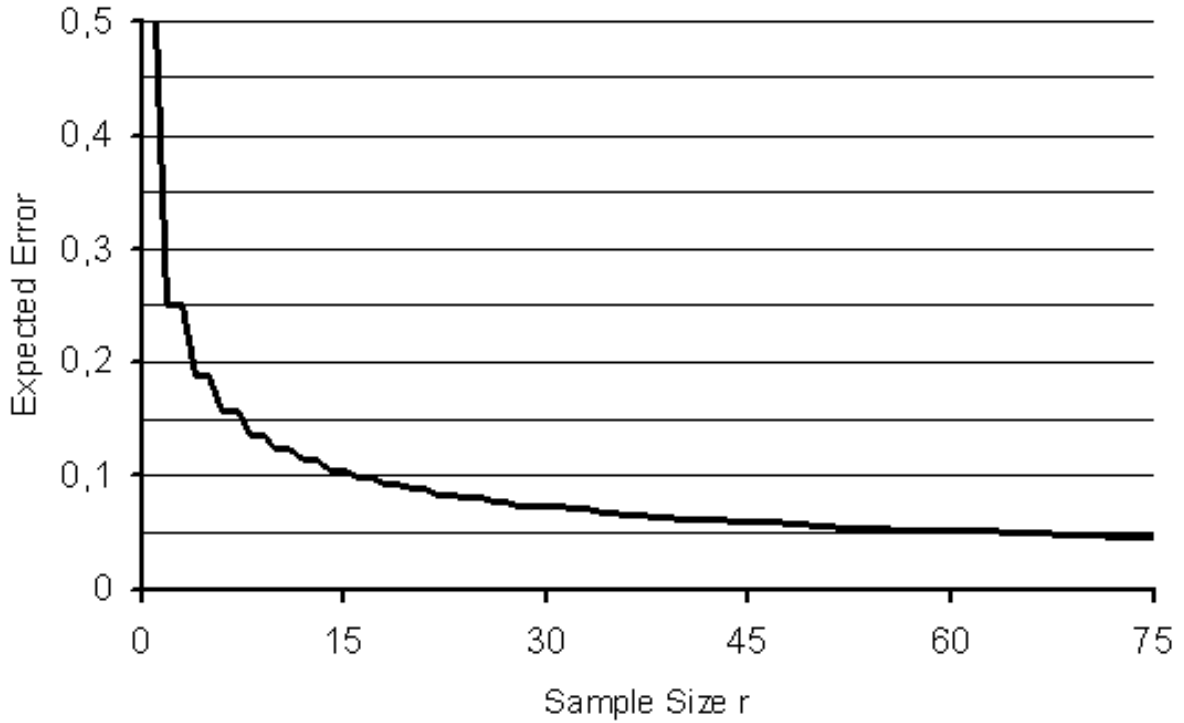

Figure 4: Expected Error over Sample Size

## 3.2 Privacy

An objective commonly opposing accuracy is privacy, and this remains the case with respect to sampled reputation scores. The question is how well does sampling protect a single feedback score. Assume an attacker is observing its average feedback score $F_B$, e.g., after each transaction. We need to determine the probability that if he observes a lower score, a negative feedback was left. This probability determines the reliability of the observation.

We conduct the following experiment. Given $n$ and $p$ we take $r$ random samples and compute $F_B$. Now we introduce two cases:

(a) a positive feedback is left

(b) a negative feedback is left.

We choose $r$ random samples again and compute $F'_B$. We define a *privacy breach* as the event, if

- in case a: $F'_B > F_B$
- in case b: $F'_B < F_B$.

We conducted this experiment for a range of values and in most cases the probability of a privacy breach was so low that we did not observe one even for very large sample sizes. Nevertheless, interesting situations arise in the borderline cases, where privacy breaches are likely.

In the remainder of this section, we assume a sampling rate $\frac{r}{n} = \frac{1}{2}$. As we have seen in Section 3.1, this rate quickly leads to accurate estimations.

The first borderline case is when there is a very low number of feedback scores (as would occur, say, with a buyer or seller just starting out). Clearly new feedback can result in a significant change of the score. For example, if a seller has one positive score and one negative score, then the next feedback will change the score $\frac{p}{n}$ by an absolute value of $\frac{1}{6}$.

In Figure 5, we depict the results of our experiments for a small number of feedback scores. The probability of a privacy breach for such small numbers of feedback scores is shown. We assume a score $\frac{p}{n} = \frac{1}{2}$ that minimizes accuracy. The probability is depicted on a logarithmic scale and from the graph we conclude an exponential decrease. Even for two feedbacks we measured only a privacy breach probability of less than 17%. For four feedbacks this probability is already less than 4%. Thus, one way to mitigate the risk of a privacy breach in this case is to simply delay the release of an agent's average feedback score until he or she has reached some reasonable threshold, of, say, 5 or 10 scores.

The second borderline case is when an agent has a high number of positive feedback scores. Clearly, negative feedback is easier to spot in a sea of positive feedback. Assuming one has a set of nearly uniformly positive feedback scores, then another positive feedback will maintain the high average or slightly increase it and a negative feedback will lower the average score if it is sampled. Therefore the probability of a privacy breach is exactly the sampling rate $\frac{r}{n}$ in case of previous scores being uniformly positive.

In Figure 6, we show the results of our experiments for the probability of a privacy breach for high feedback scores. We assume $n = 100$ feedback ratings have been left. The probability is depicted on a logarithmic scale for $p = 95$ to $p = 100$ positive ratings among this set of scores. From the graph, we conclude an exponential decrease of the probability of privacy breach with decreasing positive ratings. Nevertheless, for 99% positive ratings, the privacy breach probability is already lower than 12%. Again, as with the borderline case of too few scores, a possible countermeasure in the case of many positive scores is to wait until a minimum number of additional ratings has been received before publishing a score. This waiting period does not have to be large, since we have seen that already a second negative feedback rating is hard to detect. We therefore only have to wait until it is likely that one negative feedback has been left.

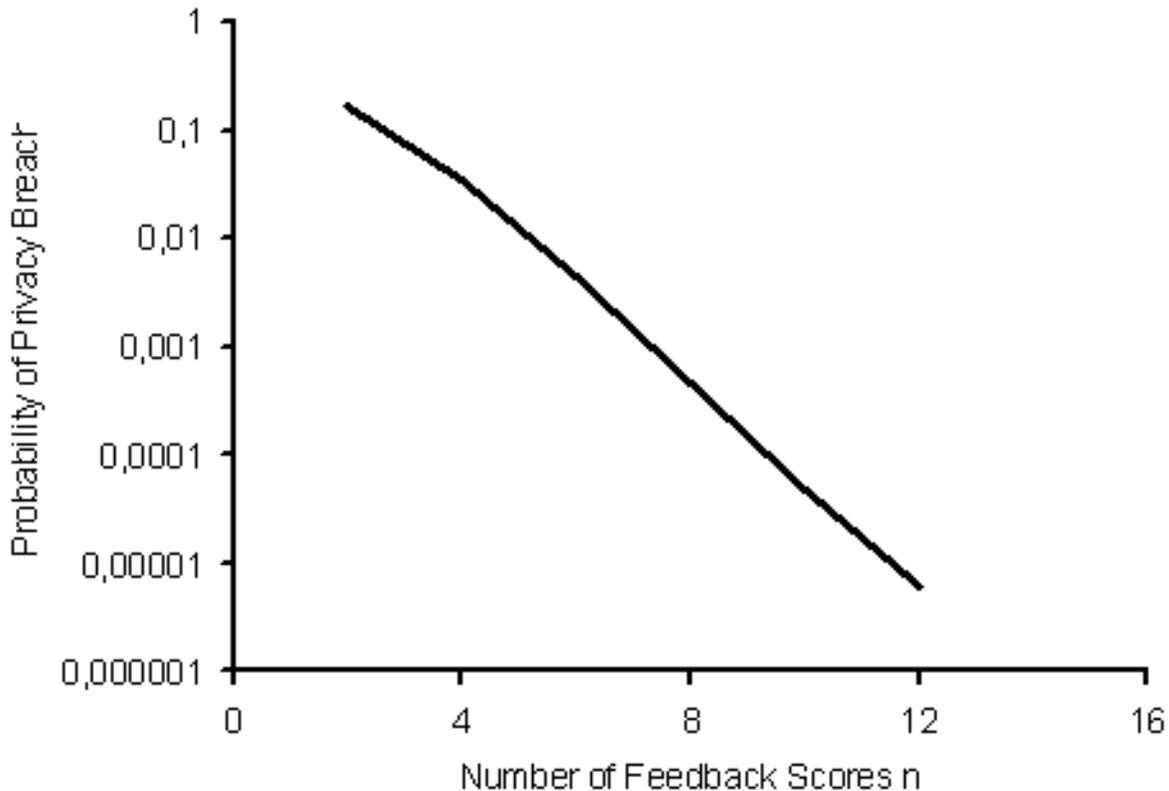

Figure 5: Probability of privacy breach for a small number of feedback scores.

We stress that these experiments rely on a single sample for each feedback score. If the feedback score is computed multiple times with different sampling sets, but identical feedback ratings, the probability of a privacy breach increases. This risk can be reduced, however, by using a sampling rate that is not too close to 1, so that it would be difficult for an adversary to know when the scores from various agents are included.

## 4 Untrusted Reputation Provider

So far we have considered methods for protecting the rater and ratee in an online auction, but the service provider aggregating the reputation scores was assumed to be fully trustworthy. Ideally, the rater and ratee should not have to fully trust the reputation provider. For instance, in [12], a scheme has been presented for distributing trust over two reputation providers $SP_1$ and $SP_2$.

The advantage of this scheme is that the first reputation provider, $SP_1$, only learns the ratings and the ratee whereas the second reputation provider only learns the rater (in addition to the reputation score). This distribution ensures privacy of the rating against a single service provider. Furthermore, using verification algorithms, rater and ratee can ensure that all, but only valid, ratings have been aggregated into the reputation score, i.e., it is not possible for the service provider to invent or suppress ratings.

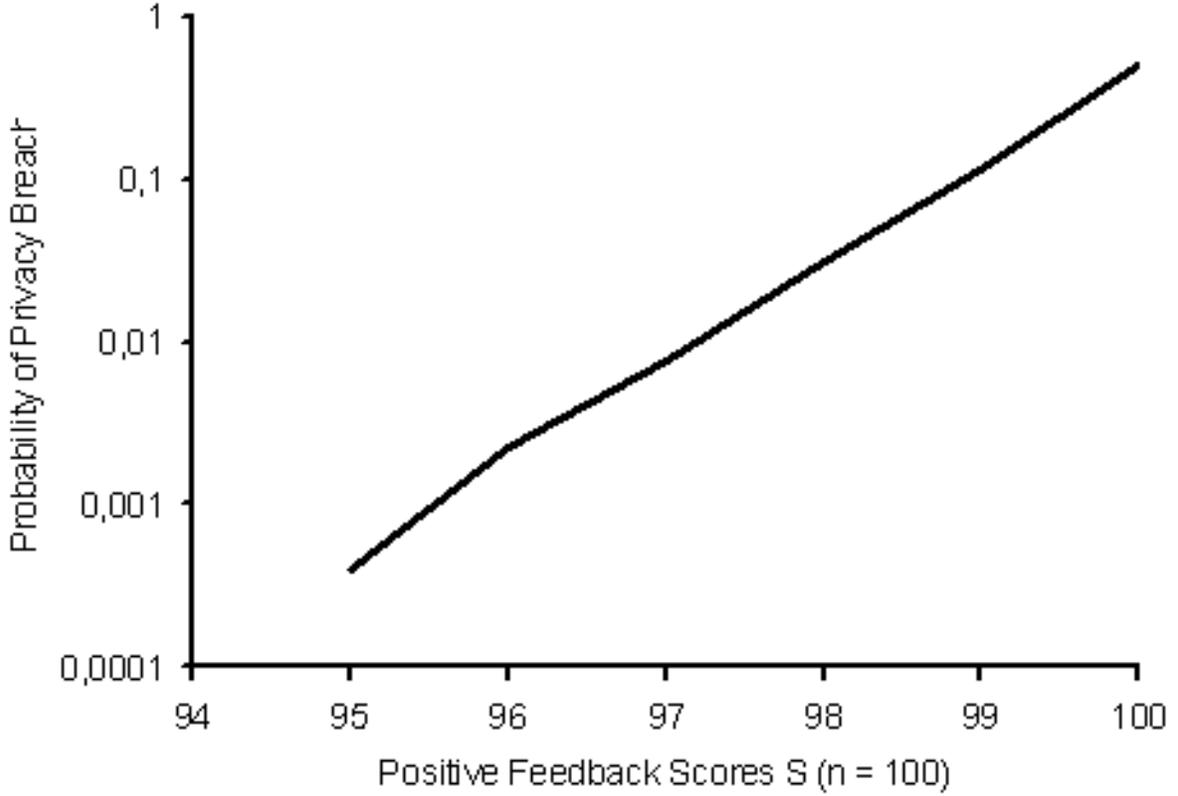

Figure 6: Probability of Privacy Breach for High Feedback Score

In the scheme from Kerschbaum [12], each transaction has a unique identifier $g^r$ and rating $z$. The first service provider publishes a public key and we denote with $E()$ the encryption using this public key.[4] The scheme consists of the following protocols and algorithms:

- Alice (ratee) and Bob (rater): $A \longleftrightarrow B$ TokenIssue() $\rightarrow$ Alice: $r$, Bob: $g^r, \ldots$[5].

- Bob and $SP_2$: $B \longleftrightarrow SP_2$ RatingSubmission() $\rightarrow SP_2$: $g^r, E(z) \ldots$.

- $SP_2$: PublishRatings() $\rightarrow g^r, E(z), \ldots$.

- Alice or Bob: VerifyRatings() $\rightarrow \top \vee \bot$

- $SP_1$: PublishReputation() $\rightarrow s = f(z, \ldots), \ldots$.

- Alice (ratee): VerifyReputation() $\rightarrow \top \vee \bot$

---

[4]We omit the identity as a subscript, since there is only one public key in the system.
[5]We omit the details for achieving security against false ratings or false aggregation.

The scheme consists of two basic mechanisms for security. For integrity, it relies on the bilinear decisional Diffie-Hellman assumption and the idea that intermediate results are hard to compute, but can be used to solve an instance of the problem. We refer the reader to [12] for details. For confidentiality, it relies on encryption. A rating $z$ is encrypted using the public key of the first reputation provider $SP_1$, such that only he will learn the rating, but he will not learn who it is from.

Homomorphic encryption is used in order to compute the reputation score on encrypted ratings. For instance, Paillier encryption [16] is an instance of homomorphic encryption that is public-key and semantically secure. Semantic security in this case means that a ciphertext is indistinguishable from any other ciphertext due to randomization. We can make this randomization explicit in our notation. A plaintext $x$ with randomization $r$ is denoted by $E(x, r)$.[6]

The randomization carries through the homomorphic property, i.e.,

$$E(x_1, r_1)E(x_2, r_2) = E(x_1 + x_2, r_1 r_2)$$

$$E(x_1, r_1)^{x_2} = E(x_1 x_2, r_1^{x_2})$$

## 4.1 Escrow

The scheme in [12] was designed for the unilateral scheme and therefore obviously does not apply to a scheme with feedback escrow.

Before extending this scheme to handle escrow, we need to make it bilateral. Therefore, for each transaction let there now be two TokenIssue() protocols. One where Alice is the ratee and obtains $r_A$ and Bob is the rater and obtains $g^{r_A}$ and another one with the roles reversed and the identifiers $r_B$ and $g^{r_B}$.

The basic idea for implementing escrow is that $SP_2$ now only publishes ratings for transactions where both parties have submitted ratings. Note that $SP_1$ cannot implement escrow, since it implies learning both parties—ratee and rater—of the transaction. Instead, $SP_2$ implements escrow learning both parties of the transactions, but not the rating $z$.

The rater submits next to $g^{r_A}$ also $g^{r_A r_B}$ to $SP_2$ in the protocol RatingSubmission(), which is a common value to both instances belonging to the same transactions. The reputation provider $SP_2$ only publishes ratings for which both ratings with $g^{r_A r_B}$ have been received. Alice and Bob can prevent $SP_2$ from suppressing ratings by signaling their submissions to each other.

## 4.2 Sampling

In order to extend this scheme further to handle sampling, we let the second reputation provider $SP_2$ select the subset of ratings. Two questions immediately arise: First, how can we ensure that he chooses the selection fairly (i.e., non-adaptive to the rating) and, second, how can we still verify the aggregation.

The idea for solving the first issue is that the reputation provider chooses the samples first and binds himself to them using a commitment. The idea for the second is to extend the zero-knowledge proof for aggregation by sampling.

Let us start with the simplest zero knowledge proof that $E(x, r)$ is an encryption $x$. The verifier (encryptor) reveals $r$ and the prover encrypts $x$ with $r$ and verifies that the ciphertexts match.

---

[6]We omit the randomization, if it is irrelevant for the exposition.

This basic idea has been used to verify that the reputation score $s$ has been correctly aggregated. The ciphertext of the reputation score $E(s,r)$ can be computed as the product of the ciphertexts of the ratings. Then the reputation provider $SP_2$ publishes $s$ and the corresponding $r$, and the ratee verifies that this corresponds to the product of the ciphertexts.

Of course, when introducing sampling, this sampling needs to be considered in the aggregation. A sampling is nothing else than a product with a bit $b \in \{0,1\}$, however. If the bit $b$ is one, the rating is selected; if the bit is zero, it is not. So we now let the second reputation provider $SP_1$ select this bit and instead of publish the rating $E(z)$, he publishes $E(bz)$.

The zero-knowledge proof of the first reputation provider $SP_1$ then remains unchanged. Furthermore, $SP_2$ can even prove that it adhered to the sampling rate by revealing the sum and randomness of the ciphertext of the sum of the sampling bits in the same way.

The remaining problems are for $SP_2$ to commit to $b$ and prove in zero-knowledge that $E_{SP_1}(bz)$ has a plaintext of the product of the committed bit and the rating.

As a commitment we let $SP_2$ publish $c = E_{SP_1}(b, r)$. In [6], there is a zero-knowledge proof (ZKP) that $c$ is a ciphertext for one-out-of-two plaintexts (which are 0 and 1 in our case). $SP_2$ publishes such a ZKP along with $c$ and keeps $b$ and $r$ for his private records. In order to make each ZKP non-interactive the first reputation provider $SP_1$, which is not involved in any such ZKP, publishes a common random string. We can then use the techniques from [3] in order to make all proofs non-interactive.

In [6], another ZKP is given for proving, for $E(\alpha, r_\alpha)$, $E(\beta, r_\beta)$, $E(\gamma, r_\gamma)$ that $\gamma = \alpha\beta$. We can use this proof for our scheme, since if $SP_2$ also publishes $E_{SP_1}(z)$ and proves that $E_{SP_1}(bz)$ is indeed a ciphertext of the product, then the privacy of the rating against $SP_2$ is still maintained, but Alice and Bob can still verify the ratings using VerifyRating().

We briefly review the ZKP from [6].
*Public Input:* $e_\alpha = E(\alpha, r_\alpha)$, $e_\beta = E(\beta, r_\beta)$, $e_\gamma = E(\gamma, r_\gamma)$
*Private Input of the Prover:* $\alpha$, $r_\alpha$, $\beta$, $r_\beta$, $\gamma$, $r_\gamma$

The prover chooses random values $d$, $r_d$, $r_{d\beta}$ and publishes $e_d = E(d, r_d)$ and $e_{d\beta} = E(d\beta, r_{d\beta})$. The prover retrieves $u$ from the common random string and publishes $v = u\alpha + d$ and $r_1 = r_\alpha^u r_d$. Finally, the prover publishes $r_2 = r_\beta^v (r_{d\beta} r_\gamma^u)^{-1}$.

The verifier also retrieves $u$ from the common random string and verifies that $e_\alpha^u e_d = E(v, r_1)$ and $e_\beta^v (e_{d\beta} e_\gamma^u)^{-1} = E(0, r_2)$.

In summary, the *ZKP* consists of:

- $E(d, r_d)$

- $E(d\beta, r_{d\beta})$

- $v = u\alpha + d, r_1 = r_\alpha^u r_d$

- $r_2 = r_\beta^v (r_{d\beta} r_\gamma^u)^{-1}$

**Lemma 1:** *The ZKP is complete and honest-verifier zero-knowledge.*

There are two differences in our scheme to the setup of [6]. First, the two ciphertexts for the factors $a$ and $b$ are distributed over two parties: Bob and $SP_2$. The reputation service provider $SP_2$ holds the sampling bit $b$ and the rater Bob the rating $z$. Second, neither of those two parties has the secret key for the encryption scheme, but that is with the first service provider $SP_1$. Fortunately, the ZKP can be computed without knowledge of the secret key, as long as the randomness for the ciphertexts is known, but that is, of course, distributed over the two parties, again.

The basic idea is to perform a secure computation for the zero-knowledge proof, i.e., $SP_2$ and Bob jointly engage in a computation of the result of the ZKP without either learning the other party's input. We can make use of the following observation: Plaintext and randomization computations in Paillier's encryption system are in similar groups. Plaintext computations are in $\mathbb{Z}_n$ and randomization computations are in $\mathbb{Z}_{n^2}^*$. Both share the modulus $n$ of secret factorization.

Fortunately, there exists a corresponding encryption scheme by Damgard and Jurik [6] operating with plaintexts in $\mathbb{Z}_{n^2}$ for each Paillier encryption scheme [16] operating with plaintexts in $\mathbb{Z}_n$. The other homomorphic properties are preserved and even the private and public keys are identical. We denote $E'(x, r)$ the encryption of plaintext $x \in \mathbb{Z}_{n^2}$ with randomization $r \in \mathbb{Z}_{n^3}^*$ in Damgard and Jurik's encryption scheme.

In the SubmitRating() protocol, Bob then not only sends $E(z, r_z)$ (his encrypted rating), but also chooses $d$, $r_d$, and $r'$. Bob publishes $E(d, r_d)$ and $v = uz + d$ and chooses $r_1 = r_z^u r_d$, i.e., we set $z = \alpha$ and $b = \beta$ in the previous proof. The random numbers $r_{d\beta}$ and $r_\gamma$ in the previous proof are (secretly) shared between Bob and $SP_2$. Bob also submits the ciphertext $E'(r_z^u r_d, r')$.

Once the the second reputation provider $SP_2$ received Bob's submission, he chooses $r_{zb}$ and publishes the multiplication of the sampling bit $b$ and the rating $z$: $E(zb, r'_{zb}) = E(z, r_z)^b E(0, r_{zb})$ ($r'_{zb} = r_z^b r_{zb}$). Then he chooses $r_{db}$ and publishes $E(db, r'_{db}) = E(d, r_d)^b E(0, r_{db})$ ($r'_{db} = r_d^b r_{db}$). If the sampling bit $b = 1$ is one, then $SP_2$ computes

$$E'(r_2^{-1}, r'') = E'(r_z^u r_d, r')^{(r_b^v)^{-1} r_{db} r_{zb}^u}$$

If the sampling bit $b = 0$ is zero, then $SP_2$ computes

$$E'(r_2^{-1}, r'') = E'((r_b^v)^{-1} r_{db} r_{zb}^u, r'')$$

Note that this only works, because of the small domain of the sampling, since we can avoid exponentiation of encrypted values.

The first reputation provider $SP_1$ decrypts and publishes (the inverse) $r_2$. During VerifyRating() Alice and Bob can now verify that $E(zb, r'_{zb})$ is indeed an encryption of the product of $E(b, r_b)$ and $E(z, r_z)$. They only need to verify that

$$E(z, r_z)^u E(d, r_d) = E(v, r_1)$$

and that

$$E(b, r_b)^v \left(E(db, r_{db}) E(z, r_z)^u\right)^{-1} = E(0, r_2)$$

Note that the first reputation provider $SP_1$ still does not learn the rater and the second reputation provider $SP_2$ still does not learn the rating. The sampling bit is known to both reputation providers, but remains entirely secret from the rater and ratee. The first service provider learns the sampling bit from the ciphertext, but he would learn it with some error from the rating anyway.

**Lemma 2:** *The protocol above is secure in the semi-honest model, i.e., neither Bob nor the second reputation $SP_2$ learn anything that cannot be derived from their input and output.*

**Proof:** We give standard simulators for the views of the parties following the seminal work of [10]. First the reputation providers view can be simulated by ciphertexts and two random numbers for $v$ and $r_1$. Both $v$ and $r_1$ are perfect secret shares where one share is only known Bob ($d$ and $r_d$). Note that ciphertexts are under the private key of $SP_1$ and therefore cannot be decrypted by either party. Bob's view can be simulated by ciphertexts and one random number for $r_2$. Also $r_2$ is a perfect secret share due to $r_{zb}$. ∎

Our scheme does not require security against malicious attackers. Instead, following the proposal from [12], we can implement a detection procedure based on signed messages. In the case of a dispute, all parties could then open their messages and the culprit would be identified.

## 5 Conclusion

In this paper, we have studied the problem of reputation feedback extortion and how it can be mitigated. We advocated the use of feedback escrow to increase the likelihood of the various parties giving honest feedback, random sampling to decrease the ability of either party tying a specific feedback score to a specific individual, and zero-knowledge proofs as a means to reduce the need for a trusted reputation provider.

Some directions for future work include an extension of differential privacy techniques to reputation scores for online auctions when the number of auction interactions is unbounded or not known in advance.

### Acknowledgments

This research was supported in part by the National Science Foundation under grants 0724806, 0847968, 0953071, 1011840.